\documentstyle[12pt,a4]{article}
\normalsize
\def\simless{\mathbin{\lower 3pt\hbox{$\rlap{\raise 5pt\hbox{$\char'074$}}
\mathchar"7218$}}}
\def\simgreat{\mathbin{\lower 3pt\hbox{$\rlap{\raise 5pt \hbox{$\char'076$}}
\mathchar"7218$}}}
\catcode`@=11

\def\beqra{\begin{eqnarray}} \def\eeqra{\end{eqnarray}}
\def\beq{\begin{equation}}      \def\eeq{\end{equation}}

\def\fo{\hbox{{1}\kern-.25em\hbox{l}}}

\def\ch{\@startsection{section}{1}{\z@}{-3ex plus-1ex minus-.2ex}%
        {2ex plus.2ex}{\large\sc}}

\def\; \lapp \;{\raisebox{-.4ex}{\rlap{$\sim$}} \raisebox{.4ex}{$<$}}

\def\con{\ifmmode \hbox{\bf*} \else{\bf*}\fi}   
\def\scon{\ifmmode \hbox{\footnotesize\rm\bf*} \else{\footnotesize\rm\bf*}\fi}

\def\0#1{\relax\ifmmode\mathaccent"7017{#1}
        \else\accent23#1\relax\fi}              

\def\ltsim{\matrix{<\cr\noalign{\vskip-7pt}\sim\cr}}
\def\gtsim{\matrix{>\cr\noalign{\vskip-7pt}\sim\cr}}

\def\eslash{\not{\hbox{\kern-2pt $E$}}}

\catcode`@=12

\begin{document}
\hoffset=0.4cm
\voffset=-1truecm
\normalsize
\pagestyle{empty}
\def\ni{{\bar {N_i}}}    \def\nj{{\bar {N_j}}}   \def\n3{{\bar {N_3}}}
\def\li{\lambda_i}    \def\lj{\lambda_j}   \def\l3{\lambda_3}
\def\hn{h^\nu}       \def\hnij{h^\nu_{ij}}

\baselineskip=5pt
\begin{flushright}
DFPD 92/A/56
\end{flushright}
\begin{flushright}
SISSA-209/92/A
\end{flushright}
\vspace{34pt}
\centerline{{\Large \bf Dynamics of the Cosmological Quark--Hadron}}
\vskip 0.3 cm
\centerline{{\Large \bf Transition
in a Matter Dominated Universe:}}
\vskip 0.39 cm
\centerline{{\Large \bf Distribution and Evolution of
Baryon Inhomogeneities}}

\vspace{44pt}
\begin{center}
{\large \bf O. Pantano}$^{\spadesuit}$ {\large \bf and}
{\large \bf A. Riotto}$^{\clubsuit}$
\end{center}
\vskip 0.6 cm
\baselineskip=3pt
\centerline{\it  $^{\spadesuit}$ Physics Department G. Galilei,
University of Padova,}
\centerline{\it via Marzolo 8, 35131 Padova, Italy.}
\vskip 0.3 cm
\centerline{\it $^{\clubsuit}$ International School for Advanced Studies,
SISSA,}
\centerline{\it via Beirut 2-4, I-34014 Trieste, Italy.}
\vspace{46pt}
\centerline{\large{\bf Abstract}}
\vskip 0.25 cm
\baselineskip=24pt
$~~~~$
We study the dynamics of the quark-hadron transition for a scenario
in which the Universe is matter dominated and a large amount of entropy
is generated by decaying particles of mass 1--10 TeV, as suggested by a
large class of superstring--inspired models. We estimate the nucleation rate
and compute the mean separation between baryon
fluctuations generated during the transition following their
evolution up to the onset of primordial nucleosyntheisis.
\pagebreak
\setcounter{page}{1}
\hoffset=0.4cm
\voffset=-1truecm
\normalsize
\baselineskip=24pt
\setcounter{page}{1}
\def\ni{{\bar {N_i}}}    \def\nj{{\bar {N_j}}}   \def\n3{{\bar {N_3}}}
\def\li{\lambda_i}    \def\lj{\lambda_j}   \def\l3{\lambda_3}
\def\g{\Gamma_{\phi}}
\def\p{\phi}
\def\hn{h^\nu}       \def\hnij{h^\nu_{ij}}
\voffset=-1 truecm
\normalsize
The cosmological Quark--Hadron Phase Transition (QHPT) has been
extensively studied in the recent years mainly in relation to
the possibility of producing baryon inhomogeneities which could
survive up to the epoch of primordial nucleosynthesis (see for a review
\cite{bp}). As first pointed out by Witten \cite{wit}, if the transition
is first order, baryon number tends to be trapped in the regions
which are still in the quark-gluon plasma state and this would lead to
an inhomogeneous distribution of baryon number at the end of the
transition. Baryon diffusion will then tend to even out these fluctuations,
but after 1 MeV the different diffusion length of neutrons and protons
could form  region which are either proton--rich and with a high--baryon
density or neutron--rich and with a low--baryon density \cite{bon}.
In these conditions the production of light elements during primordial
nucleosynthesis can be very different from that of the homogeneous case.

It is generally assumed that the QHPT takes place in a radiation dominated
Universe. This turns out to be a perfectly justifiable way to proceed in
the context of many gauge theories, including several standard grand
unified theories.

In this paper we focus on a different situation, where the energy density of
the Universe is dominated by a massive species which is still decaying
when the QHPT occurs. Even if
the results reported here can be regarded of more general validity,
a natural framework for this scenario is provided
by a large class of superstring--motivated gauge models which posses a
superheavy intermediate scale $M$, where $M$ must be larger than
$10^{15}$ GeV in order that the proton not
decay too quickly. These models can arise through the compactification
of the \mbox{$E_{8}\otimes E_{8}$} heterotic string theory \cite{green} in
a suitable Calabi--Yau space \cite{candelas}. Symmetry breaking gives rise
to four dimensional gauge groups of rank five or six \cite{witten}, which
contain as a subgroup the standard \mbox{$SU(3)_{C}\otimes SU(2)_{L}\otimes
U(1)_{Y}$}. A peculiar feature of gauge models with a superheavy
intermediate scale $M$ is the fact that  the scalar field(s) (collectively
denoted by $\phi$) responsible for $M$ acquires its vacuum expectation
value (VEV) when the temperature of the Universe approaches the
\mbox{(1--10)
TeV scale} \cite{laz}. After reaching the minimum of its potential at $M$,
the field $\phi$ starts to oscillate with a frequency which is typically
of the order of a TeV. The Universe enters a $\phi$ dominated phase
which behaves differently from the usual radiation dominated era.
We examine here which kind of implications this would have for the dynamics
of the QHPT and the generation and evolution of baryon inhomogeneities.
The key point is that, in the presence of
a superheavy intermediate scale, the QHPT can take place when the Universe
is still $\phi$ dominated.

In this paper we make the plausible assumption \cite{turner} that the $\phi$
particles disappear according to the usual exponential law. In this case
the $\phi$ particle decay does not reheat the Universe \cite{turner}.
Instead, due to the entropy increase from the decay, the temperature $T$ of
the Universe falls down more slowly (\mbox{$T\propto R^{-3/8}$} rather than
\mbox{$T\propto
R^{-1}$}, $R$ being the cosmic scale factor). As it was shown, this fact can
have far reaching consequences for axions, since the axion decay constant can
be substantially larger than $10^{12}$ GeV without causing any conflict
with cosmology \cite{axion,nucleo}.

The final temperature $T_{RH}$ is reached when the decay is over, which
occurs by the time $\g^{-1}$, where $\g$ is the decay width of the
$\phi$ particles. Therefore, the decay width $\g$ is an essential parameter
in calculating the final temperature $T_{RH}$, after which the Universe
re--enters the usual radiation dominated era.

The $\phi$ particle must decay into light particles since its mass
$M_{\phi}$ is ${\cal O}(1)$ TeV and the fields which have direct couplings
to $\phi$ become very heavy due to the VEV of $\p$, which is of
order of $M$. The effective couplings of $\p$ to the light particles through
intermediate states of mass ${\cal O}(M)$ are suppressed by a factor
$(M_{\p}/M)$,
hence the decay rate is given by \cite{yam}
\begin{equation}
\g=\kappa\frac{M_{\p}^{3}}{M^{2}},
\end{equation}
where $\kappa$ includes the effects due to field couplings,
number of decay channels, and phase space density. We expect $\kappa$
to be of order one.

Assuming that the $\p$ decay products are rapidly thermalized, and
that $R\propto t^{2/3}$ as in a matter dominated Universe, the
energy density in radiation $\rho_{R}$ is given by \cite{turner}
\beq
\rho_{R}=\rho_{R 0}\left(\frac{t}{t_{0}}\right)^{-8/3}+
\rho_{\phi 0}\left(\frac{t}{t_{0}}\right)^{-8/3}\int_{t_{0}}^{t}
\left( \frac{t^{\prime}}{t_{0}}\right)^{2/3}{\rm e}^{-\g t^{\prime}}
d\left(\g t^{\prime}\right),
\eeq
where $\rho_{R 0}$ and $\rho_{\p 0}$ are the energy density in
radiation and $\p$ particles, respectively, just after the phase transition
for the $\p$ field and $t_{0}$ is the cosmic time when the intermediate
scale phase transition takes place. The first term in eq. (2)
represents the "old" radiation  while the second term the "new" one due
to the decays of $\phi$'s.

The "new" radiation becomes the dominant component in eq. (2) for times
larger than
\beq
t^{*}\simeq
\left(\frac{5\rho_{R 0}}{ 3 \g t_{0}\rho_{\p 0}}\right)^{3/5}t_{0}.
\eeq
For intermediate scales $M$ in superstring models, $\rho_{\p 0}\sim M_{S}^{2}
M^{2}$ (where $M_{S}\sim$1 TeV is the supersymmetry breaking scale
in the
visible sector) and \mbox{$\rho_{R 0}\sim T_{0}^{4}$}
($T_{0}\simeq 1 - 10$ TeV).
Then $t^{*}$ is smaller than the time at which the QHPT occurs.

Keeping only the dominant "new" radiation term in eq. (2), $\rho_{R}$ reads
\cite{axion}
\beq
\rho_{R}=\frac{3}{5}\rho_{\p}\g t\left[1+\frac{3}{8}\g t +\frac{9}{88}\left(\g
t\right)^{
2} +\frac{27}{1232}\left(\g t\right)^{3}+....\right],
\eeq
where
\beq
\rho_{\p}=\rho_{\p 0}\left(\frac{t}{t_{0}}\right)^{-2}{\rm e}^{-\g t}
\eeq
is the energy density in the $\p$ particles at time $t$.

If we denote by $\gr$ the number of effective degrees of freedom
contributing to $\rho_{R}$ at the temperature $T$, from eq. (4) one finds
\beq
\frac{1}{3}\pi^{3}\gr T^{4} t=\g M_{P}^{2}\left[1+\frac{3}{8}\g t +
\frac{9}{88}\left(\g t\right)^{2} + \frac{27}{1232}\left(\g t\right)^{3}+
....\right],
\eeq
where $M_{P}\simeq 1.2\times 10^{19}$ GeV is the Planck mass. Note that
the time--temperature relation (6) depends only on $\g$.

For $\g t\ltsim 10^{-1}$, eq. (6) is well approximated by
\beq
\frac{1}{3}\pi^{3}\gr T^{4} t\simeq \g M_{P}^{2}.
\eeq
For $\g t=1$ we obtain the relation \cite{turner,axion}.
\beq
T_{RH}^{4}\simeq \frac{4.15}{\pi^{3}}\frac{\g^{2}M_{P}^{2}}{g_{*}
\left(T_{RH}\right)}.
\eeq
A lower bound on $T_{RH}$ of the order of $1$ MeV  is given in ref.
\cite{nucleo}. The point is that there are no satisfactory
nucleosynthesis scenarios in which the Universe is
matter dominated after $T=1$ MeV. The Universe expands more
quickly when it is matter dominated than when is radiation dominated, which
causes the transitions between neutrons and protons to decouple
earlier and allows less time for neutron decay. Both effects lead to a surfeit
of $^{4}$He. Lowering the baryon to photon ratio may postpone the breaking of
the deuterium bottleneck allowing for acceptable $^{4}$He production, but
it also results in an overabundance of $^{3}$He or D. Moreover,
if hadronic channels are open for $\p$ decays, the injected hadrons may cause
$p\leftrightarrow n$ transitions, thus affecting the primordial production
of $^{4}$He. These considerations lead to the upper bound
$\g^{-1}\ltsim 0.03$ s which is translated, according to eq. (8),
into the lower bound $T_{RH}\gtsim 6$ MeV.

If we require that the QHPT takes place when the Universe is still
$\p$ dominated, that is $T_{RH}\ltsim T_{c}\simeq (150 - 200)$ MeV, where
$T_{c}$ is the coexistence temperature for the quark and hadron phases,
a lower bound $\g^{-1}\gtsim 3.4\times 10^{-5}$ s is found. However, we
want to stress that $T_{RH}$ naturally lies in the range
\mbox{${\cal O}(1 - 100)$
MeV} and no fine--tuning on $M$ and/or $M_{\p}$ is necessary.

We assume here that the Q--H transition is a first order phase transition
and proceeds in the following way. As the Universe cools below the
the transition temperature $T_c$, it remains in the quark phase and
bubbles of the new hadron phase are nucleated within the slightly
supercooled plasma. Each bubble then starts to grow, restrained at first
by surface tension, with the phase interface behaving as a subsonic
deflagration front and pushing a compression wave out ahead of
it \cite{miller}.
Subsequently, compression waves from adjacent bubbles meet, producing a
region of disturbed fluid, and then the bubbles themselves meet and
coalesce giving rise to disconnected quark regions which proceed to shrink.

The rate ${\cal P}$ at which bubbles of hadron phase
are formed per unit volume is given by \cite{kaj,fuller}
\beq
{\cal P}(\eta)\approx C T_{c}^{4}\, {\rm exp}\left(
\frac{-16\pi\sigma^{3}}{3 T_{c} L^{2}\eta^{2}}\right),
\eeq
where $L$ is the latent heat per unit volume released during the QHPT,
$\sigma$ is the surface tension associated to the phase boundary,
$C$ is a coefficient of order unity, and $\eta$ the supercooling
parameter defined as
\beq
\eta\equiv\frac{T_{c}-T}{T_{c}}.
\eeq
Note that ${\cal P}=0$ when $T=T_{c}$ and then grows exponentially for
$T<T_{c}$.

The fraction $f(t)$ of the Universe which at the time $t$ is still
unaffected by the transition process is

\beq
f(t)={\rm exp}\left[-\int^{t}_{t_{c}}\,dt^{\prime}
f(t^{\prime}){\cal P}(t^{\prime})
\frac{4\pi}{3}v_{s}^{3}\left(\frac{T^{\prime}}{T}\right)^{8}
\left(t-t^{\prime}\right)^{3}\right],
\eeq
where $t_{c}$ is the time when the Universe first cools through $T=T_{c}$ and
$v_{s}\approx 3^{-1/2}$ is the velocity of weak shock waves driven by
nucleated bubbles of hadronic material expanding into the quark--gluon plasma.
No further nucleation is expected to occur in regions which have been
already traversed by a shock wave. Notice that in eq. (11)
$\left(T^{\prime}/T\right)$  is elevated to the eighth power, instead of the
third power as it would be in a radiation dominated Universe \cite{kaj},
since $R\propto T^{-3/8}$ in our scenario.

We now express ${\cal P}$ as a function of time by expanding about
$t=t_{f}$ \cite{al}.
This gives
\beq
{\rm ln}{\cal P}(T)\simeq{\rm ln}\,{\cal P}\left(T_{f}\right)+
\left(\frac{d\, {\rm ln}{\cal P}}{d T}\right)_{T_{f}}
\left(\frac{d T}{d t}\right)_{t_{f}}\left(t-t_{f}\right).
\eeq
The relation between the age of the Universe $t$ and the temperature $T$
can be easily read from eq. (7), with $\gr=51.25$.

{}From eqs. (7) and (12) we have
\beq
{\cal P}(t)\simeq {\cal P}\left(t_{f}\right){\rm exp}\left[
-\alpha_{\p}\left(t-t_{f}\right)\right],
\eeq
where
\beq
\alpha_{\p}=\frac{8\pi^{4}}{9}g_{*}\left(T_{f}\right)\frac{1}
{M_{P}^{2}}\frac{\sigma^{3}T_{f}^{5}T_{c}}{L^{2}\left(T_{c}-
T_{f}\right)^{3}\g}.
\eeq
Requiring that all of the Universe be reheated at $t=t_{f}$, so that
$f\left(t_{f}\right)=0$, and following ref. \cite{al}, one can show that
the mean separation $d_{\p}$ between nucleation sites is expected to be
\beq
d_{\p}=4.38\,\frac{v_{s}}{\alpha_{\p}}.
\eeq
Eq. (15) is not necessarily linked to the spectrum of the baryon number
fluctuation separations since the QHPT has only begun and, for sufficiently
small nucleation distances, bubble coalescence might cause a rearrangement
of the medium and an increase in the distance between shrinking quark
regions where baryon concentration occurs. We consider here the regime
in which it is correct to assume a duality between nucleation
sites and baryon concentration sites; thus, the same spectrum is taken to
describe both \cite{wit,al}.

During the QHPT, the scale factor of the Universe expands by a factor
$\beta_{\p}$. Because of our assumption of a duality between nucleation sites
and baryon concentration sites, the mean separation between baryon number
fluctuations at $t=t_{h}$, the time at which the Universe
is firstly completely hadronic, is given by
\beq
l_{\p}=4.38\,\beta_{\p}\,\frac{v_{s}}{\alpha_{\p}}.
\eeq
To estimate the factor $\beta_{\p}$, we suppose that the entropy per
comoving volume produced by the decay of $\p$ particles is rapidly thermalized
between relativistic particles, so that it reads
\beq
S=\frac{2\pi^{2}}{45}\gr T^{3}R^{3}.
\eeq
On the other hand, we can express the entropy per comoving volume as the
sum of two contributions, the "old entropy" and the "new" one due to
the decays of $\p$'s \cite{turner}
\beq
S=\left[S_{0}^{4/3}+\frac{4}{3}\rho_{\p 0}R_{0}^{4}
\int_{0}^{t}\left(\frac{2\pi^{2}}{45}g_{*}\left(T^{\prime}\right)\right)^{1/3}
\left(\frac{t^{\prime}}{t_{0}}\right)^{2/3}{\rm e}^{-\g t^{\prime}}
d\left(\g t^{\prime}\right)\right]^{3/4}.
\eeq
The second term on the right--hand side of eq. (18) is dominating for
times $t\gtsim t^{*}$ so that $S_{0}^{4/3}$ can be safely neglected
for temperatures ${\cal O}\left(10^{2}\right)$ MeV.

If we denote by the subscript "$-$" ("$+$") the quantities evaluated
at times immediately before (after) the beginning (completion) of the
QHPT, and assume that the Universe experiences a period of constant
temperature during the coexistence of the quark and hadron phases,
it is easy to derive that
\beq
\beta_{\p}\equiv\left(\frac{R_{+}}{R_{-}}\right)\approx
\left[\frac{g_{*}\left(T_{-}\right)}{g_{*}\left(T_{+}\right)}\right]^{8/9}
\left[\frac{\langle
g_{*}^{1/4}\left(T_{+}\right)\rangle}{\langle g_{*}^{1/4}
\left(T_{-}\right)\rangle}\right]^{8/9}.
\eeq
In the above expression $\langle g_{*}^{1/4}(T)\rangle$ means a suitable
weighted average of $g_{*}^{1/4}(T)$ in the time interval
$\left[t_{0},t\right]$
and the ratio $\left(\langle
g_{*}^{1/4}\left(T_{+}\right)\rangle/\langle g_{*}^{1/4}
\left(T_{-}\right)\rangle\right)$ can be safely set equal to unity.

The duration of the QHPT is now simply determined by the presence of the
energy--dominating $\p$ particles rather than by the details of the
transition itself as in the Standard Model (SM), where one has to solve
in a self--consistent way the Friedmann equations
to calculate the effective duration of the transition \cite{bp}.
{}From eq. (19) we have
\beq
\Delta t=\left[\left(\frac{t_{+}}{t_{-}}\right)-1\right] t_{-}\approx
\left\{\left[\frac{g_{*}\left(T_{-}\right)}{g_{*}\left(T_{+}\right)}\right]
^{4/3}-1\right\} t_{-}\approx 3.3 \,t_{-},
\eeq
where we have made use of the fact that $R$ scales as $t^{2/3}$, $g_{*}
\left(T_{-}\right)=51.25$ and $g_{*}\left(T_{+}\right)=14.25$. In the low
temperature phase we have considered only the contribution of
light leptons and photons since the contribution of pions to the effective
number of degree of freedom is negligible once that mass and volume
corrections are taken into account \cite{bp}.
{}From equation (20) we see that the length of the QHPT
is of order of the Hubble time, while in SM it is estimated to be
$10^{-2}$ the corresponding Hubble time. Such behaviour is essentially
related to the faster expansion of the scale factor which delays the
completion of the transition.

We are now ready to compare the distribution of baryon inhomogeneities
obtained in our model to those obtained assuming
SM scenario.

We can borrow the expression for $\alpha_{SM}$ from refs. \cite{fuller} and
\cite{al}
\beq
\alpha_{SM}=\frac{32\pi^{2}}{9}\left[\frac{4\pi}{5}g_{*}\left(T_{f}\right)
\right]^{1/2}\frac{1}{M_{P}}\frac{\sigma^{3}T_{f}^{3}T_{c}}{
L^{2}\left(T_{c}-T_{f}\right)^{3}}.
\eeq
{}From eqs. (14) and (21) we derive that
\begin{eqnarray}
\frac{\alpha_{\p}}{\alpha_{SM}}&=& \frac{\pi^{3/2}}{8}\sqrt{5 g_{*}
\left(T_{f}\right)}\frac{T_{f}^{5}}{\g M_{P}}\nonumber \\
&\approx&\sqrt{5 g_{*}\left(T_{f}\right)}\, \frac{
2.9\times 10^{-21}\,\mbox{GeV}}{\g}\left(\frac{T_{f}}{150\, \mbox{MeV}}
\right)^{2}.
\end{eqnarray}
Notice that the ratio $\alpha_{\p}/\alpha_{SM}$ does not
depend on $\sigma$ and $L$, but only on $T_{f}$ and $\g$.

The value of $\beta_{SM}$ can be approximated by
$\left[g_{*}\left(T_{-}\right)/g_{*}\left(T_{+}\right)\right]^{1/3}$ \cite{kaj}
so that
\beq
\frac{\beta_{\p}}{\beta_{SM}}\simeq
\left[\frac{g_{*}\left(T_{-}\right)}{g_{*}\left(T_{+}\right)}\right]
^{5/9}\simeq 2.03.
\eeq
We can finally express the ratio of the mean separation of baryon number
fluctuations in the two models at $t=t_{h}$ as
\begin{eqnarray}
\frac{l_{\p}}{l_{SM}}&=&\frac{\beta_{\p}}{\beta_{SM}}
\frac{\alpha_{SM}}{\alpha_{\p}}\nonumber \\
&\simeq& \frac{1}{\sqrt{g_{*}\left(T_{f}\right)}}\,\frac{\g}{
1.58\times 10^{-21}\,\mbox{GeV}}\left(\frac{T_{f}}{150\, \mbox{MeV}}
\right)^{-2}.
\end{eqnarray}
As stated above, we are interested in values of $T_{RH}$ in the range
$(6\ltsim T_{RH}\ltsim 150)$ MeV. From eqs. (8) and (24) we see
that $(l_{\p}/l_{SM})$ lies in the range
\beq
2.6\times 10^{-3}\ltsim \frac{l_{\p}}{l_{SM}}\ltsim 1.9.
\eeq
The behaviour of $l_{\p}/l_{SM}$ for the allowed range of $T_{RH}$ is
shown in Fig. 1.
The presence of the energy--dominating $\p$ particle can drastically
change quantitatively the spectrum of nucleation site separations,
expecially for a long matter dominated period, i.e. for small values of
$\g$. The reason for such a behaviour depends on the fact that the relation
between the time and the temperature entering the expression for $\alpha_{\p}$,
is now rather different from the standard one.
The mean distance between baryon
inhomogeneities after the completion of the QHPT is predicted to be sensibly
smaller than in SM if $\p$ particles give the major contribution to the energy
density of the Universe up to relatively small temperatures, $T={\cal O}(1)$
MeV. There is already a difficulty \cite{dif} in getting a long enough
distance scale from the quark--hadron transition in SM to let baryon
inhomogeneities survive up to primordial nucleosynthesis and the
modifications presented in this paper seem to be in the direction to make
it more difficult. On the other hand, the requirement that nonlinear
perturbations in baryon density appear larger than the proton diffusion
length, but smaller than the neutron diffusion length, must be satisfied
at the onset of nucleosynthesis in order to favour the formation
of high--baryon--density proton--rich regions and low--baryon--density
neutron--rich regions from which the light--element yields can
significantly differ from those of the standard homogeneous big bang. For
such a reason, we now proceed in estimating the mean distance between
perturbations in baryon density at $T_{6}=6$ MeV. We remind the reader
that we impose the lower bound $T_{RH}\gtsim 6$ MeV and, consequently,
after $T_{6}$ the Universe is radiation dominated and expands as
in SM scenario.

Since $R$ scales as $t^{2/3}$, from relation (7) it is straightforward
to derive that
\beq
d_{\p}\left(T_{6}\right)=l_{\p}\left(\frac{T_{+}}{T_{RH}}\right)^{8/3}
\left(\frac{T_{RH}}{T_{6}}\right)
\left[\frac{g_{*}\left(T_{+}\right)}{g_{*}\left(T_{RH}\right)}\right]
^{2/3}\left[\frac{g_{*}\left(T_{RH}\right)}{g_{*}\left(T_{6}\right)}\right]
^{1/3}.
\eeq
Note that, in the extreme case of $T_{RH}=T_{+}$,
\beq
d_{\p}\left(T_{6}\right)=l_{\p}\left(\frac{T_{+}}{T_{6}}\right)
\left[\frac{g_{*}\left(T_{RH}\right)}{g_{*}\left(T_{+}\right)}\right]^{1/3},
\eeq
that is, relation (26) reduces to the usual relation (27) which must be used
if the  radiation dominates just after the completion of the QHPT.

The ratio between the physical distances of baryon inhomogeneities in our
scenario and in SM, at $T_{weak}\sim 1$ MeV, reads
\beq
\frac{d_{\p}\left(T_{weak}\right)}{d_{SM}\left(T_{weak}\right)}
= \frac{l_{\p}}{l_{SM}}\left(\frac{T_{+}}{T_{RH}}\right)^{5/3}
\left[\frac{g_{*}\left(T_{+}\right)}{g_{*}\left(T_{RH}\right)}\right]^{1/3},
\eeq
which reduces to $(l_{\p}/l_{SM})$, as expected, if $T_{+}=T_{RH}$. The
behaviour of $d_{\p}/d_{SM}$ as function of $T_{RH}$ is illustrated
in Fig. 2. Since the evolution of the Universe after $T_{weak}$
in our scenario is the same as in SM, eq. (28) represents also the ratio
between the distances of baryon inhomogeneities in the two models
at the beginning of nucleosynthesis.

As pointed out above, upper and lower limits to scales $d_{\p}\left(T_{weak}
\right)$ on which a perturbation in baryon density can affect nucleosynthesis
through \mbox{neutron--proton} segregation are determined by diffusion lengths
of neutrons and protons at the onset of nucleosynthesis. At high temperatures
the diffusion length of neutrons and protons are equal because they
intertrasmute rapidly through weak interactions.
After weak interactions freeze out at  $T\sim T_{weak}$, nucleons retain
their identity as protons
and neutrons and diffusive segregation occur.
Coulomb collisions between protons and electrons (or positrons) give
a proton transport cross section roughly equal to the Thompson cross
section. Neutrons scatter electrons with a cross section $\sim 10^{-30}$
cm$^{2}$ because of their magnetic moment and scatter protons with a
cross section $\sim 10^{-23}$ cm$^{2}$, so that the mean free path of
neutrons is about $10^{6}$ times that of the protons.

Nucleons diffusion is described by the equation \cite{diff}
\beq
\frac{\partial n}{\partial t^{\prime}}=D_{n}\nabla^{2} n,
\eeq
where $t^{\prime}$ is a time coordinate related to the time $t$ by
\beq
dt^{\prime}=X_{n}dt.
\eeq
The factor $X_{n}(T)=\left(1+{\rm e}^{Q/T}\right)^{-1}$
represents the fraction of time spent by a nucleon as a
neutron and takes account of the fact
that all the diffusion occurs when a nucleon is a neutron.
Here $Q=$ 1.29 MeV is
the neutron--proton mass difference.

In an expanding Universe the diffusion coefficient depends on
the temperature and the baryon density and it becomes a function of time
$D_{n}(t^{\prime})$. Following ref. \cite{diff}, we can write
\beq
D_{n}(t^{\prime})=D_{0}f(t^{\prime}),
\eeq
where $D_{0}$ is a constant and $f(t^{\prime})$ is a dimensionless function
of time. Defining a new time--coordinate $u(t^{\prime})$ by
\beq
du=f(t^{\prime})dt^{\prime},
\eeq
the diffusion equation is transformed into
\beq
\frac{\partial n}{\partial u}=D_{0}\nabla^{2} n.
\eeq
After a time $t$ the rms distance diffused by  neutrons in an expanding
Universe is
\beq
d_{n}=\left[6 D_{0} u(t^{\prime})\right]^{1/2}.
\eeq
As said above, we assume that weak interactions are in equilibrium
down to a temperature $T_{weak}\sim 1$ MeV, so that the diffusion
lengths are computed from eq. (34), with $u(t^{\prime})$
defined by eq. (32) if
the weak interactions have not yet decoupled, {\it i.e.} $T>T_{weak}$, or by
the same equations, with $t\equiv t^{\prime}$ for $T<T_{weak}$.

Neutrons are scattered by electrons and positrons or by protons, so
that the neutron diffusion coefficient $D_{n}$ is given by
\begin{equation}
D_{n}^{-1}=D_{ne}^{-1}+D_{np}^{-1},
\end{equation}
where $D_{ne}$ is the neutron diffusion coefficient appropriate for
electron scattering only and $D_{np}$ is the diffusion coefficient for
neutron--proton scattering. We use the non--relativistic
expressions for $D_{ne}$ and $D_{np}$ from ref. \cite{diff}
\begin{eqnarray}
D_{ne}&=&\frac{3\pi^{2}}{16 m_{e}^{3}}\frac{1}{\sigma_{t}}\frac{
z{\rm e}^{z}}{1+3/z+3/z^{2}},\nonumber\\
D_{np}&=&\sqrt{\frac{T}{3 m_{n}}}\frac{1}{n_{p}\sigma_{np}},
\end{eqnarray}
where $z=m_{e}c^{2}/T$, $\sigma_{t}$ is the scattering cross section of
neutrons and electrons due to the interactions of their magnetic
moments, $n_{p}$ is the number proton density and $\sigma_{np}$ is the
$n$-$p$ cross section. The relativistic corrections to $D_{ne}$ and
$D_{np}$ have been recently taken into account in ref. \cite{lun} and we
expect they do not change quantitatively our conclusions, since they
mainly affect the diffusion coefficients near the nucleosynthesis
temperature.

Whereas no change for $D_{ne}$ is present in our scenario with respect
to SM value, the diffusion coefficient for $n$-$p$ scattering can be
drastically changed due to its dependence on the proton number density
$n_{p}$ and for $T>T_{RH}$ is given by
\beq
D_{np}(T)=\frac{45}{2\sqrt{3}}\frac{g_{*}(T_{RH})T_{RH}^{5}}
{m_{n}^{1/2}\pi^{2}\eta_{s}(T_{RH})\left(1-X_{n}\right)}\frac{T^{1/2}}{
g_{*}^{2}(T) T^{8}\sigma_{np}},
\eeq
where $\eta_{s}=n/s$.

The point is that the large amount of entropy released when the field
$\phi$ oscillates around its minimum decreases the value of $\eta_{s}$.
A value of $\eta_s$ at the nucleosynthesis epoch consistent with
observed light elements abundances corresponds to a value of
$n_{p}$ for $T>T_{RH}$ larger than its SM value. This  leads to
much smaller values of $D_{np}$ with respect to to SM case.
This effect is obviously exalted when $T_{RH}$ is close to $T_{6}$,
{\it i.e.} when the Universe experiences a long matter--dominated period.
In Fig. 3 we compare the behaviour of $D_{np}$ and $D_{ne}$
for a radiation--dominated Universe and for a $\p$--dominated Universe
with $T_{RH}=T_6$.
As an example, we have required that $\eta_{s}$ gets a value at the
nucleosynthesis such
that $\Omega_{b}h_{50}^{2}=1$, $\Omega_{b}$ being the proton to the
critical density ratio and $h_{50}$ the value of the Hubble parameter in
units of 50 Km sec$^{-1}$ Mpc$^{-1}$.

In our scenario the value of $D_{np}$ is lower than in SM one
for $T_{RH}\simless T\simless T_{c}$ and can give
the largest contribution to $D_{n}$ for low values of $T_{RH}$.
This situation is different in SM where it is $D_{ne}$ that
determines  the neutron diffusion distance before the onset of the
nucleosynthesis.
The fact that the diffusion coefficient can be much smaller than in SM
seems to suggest that in the scenario considered here it might be more
difficult to smear out baryon inhomogeneities before the onset of
nucleosynthesis. However, the faster expansion rate of the Universe
compensates the smaller diffusion rate so that the physically relevant
diffusion length of neutrons at 1 MeV is essentially the same as in
SM case, as illustrated by Fig. 4. The same is true at the beginning of
the nucleosynthesis since, as we have already said, the evolution
of the Universe after 1 MeV is equal to standard one.

In conclusion, we have shown that, if the
QHPT takes place when the Universe is dominated by a massive species
whose decay releases a large amount of entropy,
as suggested by a large class of superstring--motivated gauge models,
the mean distance between baryon inhomogeneities
at the end of the QHPT may be much smaller than in SM case.
In addition, the neutron diffusion coefficient can receive its major
contribution from the neutron--proton scattering rather than from the
neutron--electron scattering and might also be considerably reduced
in the scenario which we consider here.
We have also estimated the mean distance between baryon inhomogeneities
and neutron diffusion distance at the beginning of primordial
nucleosynthesis. We have found that they are not appreciably
different from those in SM mainly because the Universe
is expanding faster. As a consequence, an inhomogeneous
nucleosynthesis remains a viable possibility
within the scenario considered in this Letter.

\vspace{1 cm}
\begin{flushleft}
{\Large{\bf Figure Captions}}
\end{flushleft}
\begin{description}
\vskip 0.3 cm
\item[Figure 1] Ratio of the distance $l_{\phi}/l_{SM}$ between
baryon inhomogeneities
in the $\phi$--dominated Universe and SM radiation dominated
Universe  at $T_{c}=150$ MeV as a function of the reheating
temperature $T_{RH}$.
\item[Figure 2] Ratio of the distance $d_{\phi}/d_{SM}$
between baryon inhomogeneities
in the $\phi$--dominated Universe and
SM radiation dominated
Universe  at $T_{weak}=1$ MeV as a function of the reheating
temperature $T_{RH}$.
\item[Figure 3] Neutron--electron diffusion coefficient $D_{ne}$
(solid line), neutron--proton diffusion coefficients $D_{np}$ in the
$\phi$--dominated Universe (dashed line) and in  SM
radiation--dominated Universe (long--dashed line) as a function of the
temperature. Reheating temperature $T_{RH}$ is set equal to 6 MeV.
\item[Figure 4] Diffusion distance of neutrons as a function of the
temperature in the $\phi$--dominated Universe (dashed line) and in
SM radiation--dominated Universe (solid line). The reheating temperature
$T_{RH}$ is set equal to 6 MeV.
\end{description}

\end{document}